\documentstyle[12pt]{article}
\begin{document}
\begin{center}
{\bf{THE SEIBERG-WITTEN MAP IN NONCOMMUTATIVE FIELD THEORY:\\
 AN ALTERNATIVE INTERPRETATION}}
\vskip 2cm
Subir Ghosh\\
\vskip 1cm
Physics and Applied Mathematics Unit,\\
Indian Statistical Institute,\\
203 B. T. Road, Calcutta 700108, \\
India.
\end{center}
\vskip 3cm
{\bf Abstract:}\\
In this article, an alternative interpretation of the Seiberg-Witten map in 
non-commutative
field theory is provided. We show that the Seiberg-Witten map can be induced 
in a geometric way, by a field  dependent co-ordinate transformation that
connects noncommutative and ordinary space-times. Furthermore, in 
continuation of  our earlier
works, it
has been demonstrated here that the above (field  dependent co-ordinate) 
transformations are
present in a gauge fixed version of the  relativistic spinning particle 
model, embedded in
the Batalin-Tyutin extended space.
We emphasize that  the space-time non-commutativity emerges naturally from 
the particle
{\it {spin}} degrees of freedom. Contrary to similarly motivated works, the 
non-commutativity
is not imposed here in an {\it{ad-hoc}} manner.
\newpage
\begin{center}
{\bf {INTRODUCTION}}
\end{center}
\vskip .5cm

 Historically the Non-Commutativite (NC)  spacetime was  introduced by Snyder 
\cite{sn} as
 a regularization to tame the short distance singularities, inherent in a 
Quantum Field
 Theory (QFT). In some sense, this extra structure might appear as
  a natural generalization of the phase space
non-commutativity  in quantum mechanics. The softening of the
divergence  is obviously because NC in spacetime can introduce a
lower bound in the continuity of spacetime, just as $\hbar$ does
in the quantum phase space. The advantage of NC as a
regularization is that the computational scheme requires very
little changes from the ordinary spacetime and in some forms of NC
\cite{sn}, (for more recent works see \cite{lor, sg3,mor}),
manifest Lorentz invariance can be maintained. {\footnote { The
ideas of \cite{sg3} is discussed in a field theoretic setting in
\cite{gh2}.}}All the same, the idea of Snyder \cite{sn} lost
popularity due to the success of renormalization techniques in
QFT. Also, now we know \cite{rev} that  noncommutativity is not of
much use as a regulating scheme.

However, in more recent times, NCQFT has matured as an area of
intense research activity \cite{rev}. It has been established by
Seiberg and Witten \cite{sw}  that the existence of
non-commutativity  in  (open) string boundaries in the presence of
a constant two-form Neveu-Schwarz field results in NC D-branes to
which the open string endpoints are attached. (Hamiltonian
formulation of the above open string boundary noncommutativity is
discussed in \cite{bbg}.) This renders QFTs living on the D-brane
noncommutative. The authors in \cite{sw} also propose an explicit
mapping between NC and ordinary space-time dynamical variables.
This celebrated map goes by the name of the  Seiberg-Witten Map
(SWM) \cite{sw}. Actually the SWM is the culmination of a much
deeper understanding between the connection of QFTs in NC and
ordinary space-time.  It has been  shown in \cite{sw} that the
appearance of NCQFT is dependent on the choice of regularization
and in fact a QFT in ordinary spacetime and an NCQFT both can
describe the same underlying QFT. At least to the lowest
non-trivial order in $\theta _{\mu\nu} $ {\footnote{The SWM for
higher orders in $\theta$ are discussed in \cite{of}.}}, the
non-commutativity parameter,
\begin{equation}
[x_{\mu},x_{\nu}]=i\theta_{\mu\nu},
\label{0}
\end{equation}
the SWM can be exploited to convert a NCQFT to its counterpart
living in ordinary space-time, in which the effects of
non-commutativity appear as local interaction terms, supplemented
by $\theta _{\mu\nu} $. In the more popular form of  NCQFT,
$\theta _{\mu\nu} $ is taken to be constant. This can lead to very
striking signatures in particle physics  phenomenology  in the
form of Lorentz symmetry breakdown, new interaction vertices etc.
\cite {phen}.

The SWM  plays a pivotal role in our understanding of the NCQFT by
directly making contact between NCQFT and QFT in ordinary
space-time. This has led to new results in the form of axial
anomaly \cite{sb} in an NC   interacting theory of fermions
coupled to gauge fields. The previous results \cite{ard} are
incomplete as they do not conform to the SWM. Furthermore, the SWM
is crucial in generalizing \cite{ss} to NC space-time,  the
duality between  Maxwell-Chern-Simons and Self-Dual theories in
2+1-dimensions \cite{ds}. In the context of bosonization of the
massive Thirring model in 2+1-dimensions \cite{f}, SWM reveals
that the resulting theory is different from  the NC self-dual
model  \cite{gh}, in contrast to ordinary space-time.

However, as it stands, the SWM  is linked exclusively to NC {\it gauge}
theory, since the original derivation of the SWM \cite{sw} hinges
on the concept of identifying gauge orbits in NC and ordinary
space-times. In the explicit form of the SWM \cite{sw}, the
non-commutativity of the {\it space-time} in which the NC gauge
field lives, is not manifest at all since the map is a relation
between the NC and ordinary gauge fields and gauge transformation
parameters, all having ordinary space-time coordinates as their
arguments.

On the other hand, possibly it would have been more
natural to consider first a map between NC and ordinary space-time
and subsequently to induce the SWM through the  change in the
space-time argument of the gauge field from the ordinary to NC
one. Precisely this type of a geometrical reformulation of the SWM
is proposed in the present work.

The paper is organized as follows: {\bf {Section II}} describes the motivation
and an outline of our ideas. The SWM is analysed in a different light in {\bf
{Section III}}.
This constitutes our alternative interpretation of the SWM. {\bf{Section IV}} 
is devoted to
a discussion on the Batalin-Tyutin extension and the ordinary and 
noncommutative spacetime
mapping. The Batalin-Tyutin extension and SWM are bridged through the process 
of gauge fixing
in {\bf{Section V}}. The paper ends with a conclusion and future prospects in 
{\bf{Section VI}}.
\vskip 1cm
\begin{center}
{\bf {II. MOTIVATION AND AN OUTLINE OF THE PRESENT WORK}}
\end{center}
\vskip .5cm

In the canonical quantization prescription, the Poisson Bracket
algebra is elevated to  quantum commutator algebra by the
replacement
$$\{A,B\}\rightarrow \frac{1}{i}[\hat A,\hat B].$$ But presence of {\it 
constraints} may
demand a modification in the Poisson Bracket algebra, leading to the Dirac 
Bracket
algebra \cite{d}, which are subsequently identified to the commutators,
$$\{A,B\}_{DB}\rightarrow \frac{1}{i}[\hat A,\hat B].$$ However, 
complications in this
formalism can arise, (in particular in case of non-linear constraints), where 
the Dirac
Bracket algebra itself becomes operator valued. To overcome this, Batalin and 
Tyutin \cite{bt}
have developed a systematic scheme in which all the physical variables are 
mapped in an
extended canonical  phase space, consisting of auxiliary degrees of freedom 
besides the
physical ones, with all of them enjoying canonical free Poisson Bracket 
algebra. In this
formalism, the ambiguity of using (operator valued) Dirac Brackets as quantum 
commutators
does not arise.

In the spinning particle model \cite{sg1} the canonical $\{x_{\mu},x_{\nu}\}=0
$ Poisson
Bracket changes to an operator valued Dirac Bracket,
\begin{equation}
\{x_{\mu},x_{\nu}\}_{DB}=-\frac{S_{\mu\nu}}{M^{2}},
\label{dir}
\end{equation}
due to the presence of constraints. In the above, the dynamical variable $S_
{\mu\nu}$
represents the spin angular momentum and $M$ is the mass of the particle. 
This forces us to
exploit the Batalin-Tyutin prescription \cite{bt}.

In a recent paper \cite{sg3}, we have constructed a mapping of the form,
\begin{equation}
\{x_\mu ,x_\nu \}=0~~,~~
x_\mu \rightarrow \hat x_\mu ~~;~~
\{\hat x_\mu ,\hat x_\nu \}=\theta _{\mu\nu},
\label{nc}
\end{equation}
which bridges the noncommutative and ordinary space-times. Note that $\hat x_
{\mu}$ lives
in the Batalin-Tyutin \cite{bt} extended space and is of the generic
form $\hat x_{\mu}=x_{\mu}+X_{\mu}$, where $X_{\mu}$ consists of physical and 
auxiliary
degrees of freedom. Explicit expressions for $X_{\mu}$ are to be found later 
\cite{sg3}.

This space-time map induces in a natural way the following map between 
noncommutative and
ordinary degrees of freedom,
\begin{equation}
\lambda (x)\rightarrow \lambda (\hat x)
\rightarrow \hat \lambda (x)~~,~~A_\mu (x)\rightarrow A_\mu (\hat x)
\rightarrow \hat A_\mu (x).
\label{2}\end{equation}
Here $\hat \lambda$ and $\hat A_\mu$ are the NC counterparts of $\lambda$ and 
$A_\mu$, the
abelian gauge transformation parameter and the gauge field respectively and 
$\hat x_\mu $
and $x_\mu $ are the NC and ordinary space-time co-ordinates.

On the other hand, there also exists the SWM \cite{sw} which interpolates 
between
noncommutative and ordinary variables,
\begin{equation}
\lambda (x)
\rightarrow  \hat \lambda (x)~~,~~A_\mu (x)\
\rightarrow \hat A_\mu (x).
\label{2a}\end{equation}

It is only logical that the above two schemes ((\ref{nc}-\ref{2}) and \ref
{2a}) can be related.
In the present work we have precisely done that. The formulation \cite{sg3}
(\ref{nc}-\ref{2}) being the more general one, we have explicitly 
demonstrated how it can be
reduced to the SWM \cite{sw}, in a particular gauge. This incidentally 
demonstrates the
correctness of the procedure. The above idea was hinted in \cite{sg3}.
\footnote {The present analysis being classical, (non)commutativity is to be 
interpreted
in the sense of Poisson or Dirac Brackets.}

In this context, let us put the present work in its proper perspective. 
Recently a
number of works have appeared with the motivation of recovering
the SWM in a geometric way, without invoking the gauge theory
principles \cite{corn}. However, the noncommutative feature of the
space-time plays no direct role in the above mentioned
re-derivations of the SWM, with non-commutativity just being postulated in an 
{\it ad hoc} way.
In the present work, we have shown how
one can construct a noncommutative sector inside an extended phase
space, in a relativistically covariant way. More importantly, we
have shown explicitly how this generalized map can be reduced to
the SWM under certain approximations. Interestingly, this extended
space is physically significant and well studied: It is the space
of the relativistic spinning particle \cite{sg3,sg1}.
 Hence it might be intuitively appealing to think that the NC space-time is
 endowed with spin degrees of freedom, as compared to the ordinary 
configuration
 space, since the spin variables directly generate the NC \cite{gh2}. The 
analogue of the
 gauge field is also  identified inside this phase space, without
any need to consider external fields. This situation is to be
contrasted with the NC arising from the background magnetic field
in the well known Landau problem \cite{rev} of a charge moving in a plane in 
the presence of
a strong, perpendicular magnetic field.
\vskip 1cm
\begin{center}
{\bf {III. SEIBERG-WITTEN MAP: AN ALTERNATIVE INTERPRETATION}}
\end{center}
\vskip .5cm

 The genesis of the SWM is the observation \cite{sw} that the
non-commutativity in string theory depends on the choice of the regularization
scheme: it appears in {\it e.g.} point-spitting regularization
whereas it does not show up in Pauli Villars  regularization. This
feature, among other things, has prompted Seiberg and Witten
\cite{sw} to suggest the  map connecting the NC gauge fields and
gauge transformation parameter to the ordinary gauge field and
gauge transformation parameter. The explicit form of the
SWM \cite{sw}, for abelian gauge group, to the
first non-trivial order in the NC parameter $\theta _{\mu\nu }$ is
the following,
$$
\hat \lambda (x)=\lambda (x)+{1\over 2}\theta ^{\mu\nu}A_\nu(x)\partial 
_\mu\lambda (x)
~,$$
\begin{equation}
 \hat A_\mu(x)=A_\mu(x)+{1\over 2}\theta ^{\sigma\nu}A_\nu(x)F_{\sigma\mu}(x)+
{1\over
2}\theta ^{\sigma\nu}A_\nu(x)\partial _\sigma A_\mu (x).
 \label{1}
\end{equation}

The above relation (\ref{1}) is an $O(\theta )$  solution of the  general map 
\cite{sw},
\begin{equation}
\hat A_\mu (A+\delta _\lambda A)=\hat A_\mu (A)+\hat\delta _{\hat\lambda}\hat 
A_\mu(A),
\label{swm}
\end{equation}
which is based on identifying gauge orbits in NC and ordinary
space-time.

First let us show that it is indeed possible to re-derive the SWM
using geometric objects. We rewrite the SWM (\ref{1})
in the following way,
\begin{equation}
\hat \lambda (x) =\lambda (x)+\frac{1}{2}\{\delta _f[\lambda (x)]-(\lambda
(x')-\lambda (x))\}=\lambda (x)+\delta _f[\lambda (x)]~,
\label{3}
\end{equation}
\begin{equation}
\hat A_\mu (x)=A_\mu (x)+\{\delta
_f[A_\mu (x)]-(A_\mu (x')-A_\mu (x))\}=A_\mu (x)+A'_\mu (x)-A_\mu (x'). \label
{4}
\end{equation}
In the above we have defined,
$$
x'_\mu =x_\mu -f_\mu ~~,~~A'_\mu (x')={{\partial x^\nu}\over {\partial
x'^\mu}}A_\nu (x)~~,~~ \lambda '(x')=\lambda (x)~,$$
\begin{equation}
 f^\mu\equiv {1\over
2}\theta ^{\mu\nu}A_\nu ~~. \label{5}
\end{equation}
Here $f^\mu $ is the field dependent space-time translation parameter and
$\delta _f$ constitutes the Lie derivative connected to $f^\mu $,
$$
\delta _f[\lambda (x)]=\lambda '(x)-\lambda (x) =-(\lambda
(x')-\lambda (x))=f^i\partial _i\lambda ,
$$
\begin{equation}
\delta _f[A_\mu (x)]=A'_\mu (x)-A_\mu (x)~.
\label{6}
\end{equation}
This shows that the NC gauge parameter $(\hat \lambda )$ and gauge
field $(\hat A^\mu)$ are derivable from the ordinary one by making
a {\it field dependent} space-time translation $f^\mu$ \cite{j}. One can check
that the NC gauge transformation of $\hat A_\mu (x)$ is correctly reproduced
by considering,
\begin{equation}
\hat\delta \hat A_\mu (x)=\delta (A_\mu(x)+{1\over 2}\theta
^{\sigma\nu}A_\nu(x)F_{\sigma\mu}(x)+{1\over
2}\theta ^{\sigma\nu}A_\nu(x)\partial _\sigma A_\mu (x))~,
 \label{7}
\end{equation}
where $\delta A_\mu(x)=\partial _\mu \lambda (x)$ is the gauge transformation
in ordinary space-time. Hence, if expressed in the form (\ref{4}), the SWM,
(at least to $O(\theta )$), can be derived in a geometrical way, without
introducing the original identification (\ref{swm}) obtained from the
viewpoint of a matching between NC and ordinary gauge invariant sectors. Also 
note that the
gauge field $A_\mu (x)$ is treated here just as an ordinary vector field, 
without invoking any
gauge theory properties. This constitutes the first part of our result.

Returning to our starting premises, are we justified in making an
identification between $\hat x_\mu $ in (\ref{nc})-(\ref{2}) and
$x'_\mu $ introduced in (\ref{3})-(\ref{5}), because this relation can 
connect NC and ordinary
space-time. Naively, a relation
of the form, $x'_\mu =x_\mu -f_\mu (x)$ can not render the
$x'$-space noncommutative, since the right hand side of the
equation apparently comprises of commuting objects only. In our subsequent 
discussion we will
show how this surmise can be made meaningful and will return to this point at 
the end.
\vskip 1cm
\begin{center}
{\bf {IV. BATALIN-TYUTIN EXTENSION: $x_{\mu}\rightarrow \tilde x_{\mu} $ 
MAPPING}}
\end{center}
\vskip .5cm

We start by considering a larger space having inherent NC. Such a space, 
which at the
same time is physically appealing, is that of the Nambu-Goto model of
relativistic spinning particle  \cite{sg1,sg3}.
 Here the situation is somewhat akin to
the open string boundary NC such that the role of Neveu-Schwarz field is 
played by here by
the spin degrees of freedom. The Lagrangian of the  model in
2+1-dimensions \cite{sg1,sg3} is,
\begin{equation}
L=[M^2u^\mu u_\mu+{{J^2}\over 2}\sigma ^{\mu\nu }\sigma _{\mu\nu }+MJ\epsilon
^{\mu\nu\lambda}u_\mu \sigma _{\nu\lambda } ]^{{1\over 2}} ~,
\label{8}
\end{equation}
\begin{equation}
u^\mu={{dx^\mu }\over {d\tau }} ~~,~~\sigma ^{\mu\nu }=
\Lambda _\rho ^{~\mu } {{d\Lambda ^{\rho\nu }}\over {d\tau
}}=-\sigma ^{\nu\mu }~~,~~\Lambda _\rho ^{~\mu }\Lambda^{\rho\nu } =\Lambda _
{~\rho
}^\mu \Lambda^{\nu \rho }=g^{\mu\nu}~~,~~g^{00}=-g^{ii}=1.
\label{88}
\end{equation}
Here $(x^\mu ~,~\Lambda ^{\mu\nu })$ is a Poincare group element
and also a set of dynamical variables of the theory.

In a Hamiltonian formulation, the conjugate momenta are,
\begin{equation}
P^\mu={{\partial L}\over {\partial u_\mu}} =L^{-1}[M^2u^\mu
+{{MJ}\over 2}\epsilon ^{\mu\nu\lambda}\sigma _{\nu\lambda }]~~,~~
S^{\mu\nu}={{\partial L}\over {\partial \sigma _{\mu\nu
}}}={{L^{-1}}\over 2}[J^2\sigma ^{\mu\nu } +{MJ}\epsilon
^{\mu\nu\lambda}u_\lambda ].
 \label{10}
\end{equation}
The Poisson algebra of the above phase space degrees of freedom are,
\begin{equation}
\{P^\mu ,x^\nu \}=g^{\mu\nu}~~,~~\{P^\mu ,P^\nu \}=0 ~~,~~\{x^\mu
,x^\nu \}=0~~,~~\{\Lambda^{0\mu},\Lambda ^{0\nu}\}=0~,
 \label{011}
\end{equation}
\begin{equation}
\{S^{\mu\nu},S^{\lambda\sigma}\}=S^{\mu\lambda}g^{\nu\sigma}-S^{\mu\sigma}
g^{\nu\lambda}+S^{\nu\sigma}g^{\mu\lambda}-S^{\nu\lambda}g^{\mu\sigma}~,~
\{\Lambda
^{0\mu},S^{\nu\sigma}\}=\Lambda ^{0\nu}g^{\mu\sigma}-\Lambda
^{0\sigma}g^{\mu\nu}~.
\label{012}
\end{equation}
The full set of constraints are,
\begin{equation}
\Psi _1\equiv P^\mu P_\mu -M^2\approx 0 ~~,~~\Psi _2\equiv S^{\mu\nu} S_
{\mu\nu}-2J^2\approx 0,
 \label{11}
\end{equation}
\begin{equation}
\Theta _1^\mu \equiv S^{\mu\nu}P_\nu~~~,~~~\Theta _2 ^\mu \equiv \Lambda
^{0\mu}-{{P^\mu }\over M}~~,~~\mu =0,~1,~2~,
 \label{12}
\end{equation}
out of which $\Psi _1$ and $\Psi _2$ give the mass and spin of the particle 
respectively.
\footnote {Note that instead of $\Psi _2$ as above, one can equivalently use
$\Psi _2\equiv \epsilon
^{\mu\nu\lambda}S_{\mu\nu}P_\lambda -MJ, $
which incidentally defines the Pauli Lubanski scalar.}
 In the Dirac constraint analysis \cite{d}, these are termed as First Class 
Constraints
 (FCC), having the property that they commute with {\it all} the constraints 
on the
 constraint surface and
 generate gauge transformations. The set $\Theta _2 ^\mu$ is put by hand \cite
{sg3}, to
 restrict
 the number of angular co-ordinates.

The non-commuting set of constraints $\Theta _\alpha ^\mu~,\alpha =1,2 $,
termed as Second Class Constraints (SCC) \cite{d}, modify the
Poisson Brackets (\ref{011}) to Dirac Brackets \cite{d}, defined
below for any two generic variables $A$ and $B$,
\begin{equation}
\{A,B\}_{DB}=\{A,B\}-\{A,\Theta _\alpha ^\mu\}
\Delta _{\mu\nu}^{\alpha \beta }\{\Theta _\beta ^\nu, B\} ,
\label{22}
\end{equation}
\begin{equation}
\{\Theta ^\mu _\alpha  ,\Theta ^\nu _\beta \}\equiv\Delta ^{\mu\nu}_
{\alpha\beta}~~,~~\alpha ,\beta =1,2~~,
~~\Delta ^{\mu\nu}_{\alpha\beta}\Delta _{\nu\lambda}^
{\beta\gamma }=\delta _\alpha ^\gamma\delta ^\mu_\lambda~.
\label{014}
\end{equation}
$\Delta ^{\mu\nu}_
{\alpha\beta}$ is non-vanishing even on the constraint surface.
The main result, relevant to us, is the
following Dirac Bracket \cite{sg1,sg3},
\begin{equation}
\{x_\mu ,x_\nu \}_{DB}=-\frac{S_{\mu\nu}}{M^2} \rightarrow \{\hat x_\mu ,\hat 
x_\nu \}=
\theta _{\mu\nu}.
\label{db}
\end{equation}
This is the non-commutativity that occurs naturally
in the spinning particle model. Our aim is to express this NC
co-ordinate $\hat x_\mu$ in the form $\hat x_\mu= x_\mu -f_\mu$, with the 
identification
between $\theta _{\mu\nu}$ and
$S_{\mu\nu}$. This is indicated in the last equality in (\ref{db}).
In the quantum theory, this will lead to the NC space-time
(\ref{nc}).

This motivates us to the Batalin-Tyutin quantization \cite{bt} of the spinning
particle \cite{sg3}. For a system of irreducible SCCs, in this formalism \cite
{bt}, the phase
space is extended by introducing additional BT variables,
$\phi
 ^\alpha _a $, obeying
\begin{equation}
\{\phi ^\alpha _\mu,\phi ^\beta _\nu\}=\omega ^{\alpha \beta}_{\mu\nu}=
-\omega ^{\beta \alpha}_{\nu\mu}~~,~~\omega ^{\alpha \beta}_{\mu\nu}=
g_{\mu\nu}\epsilon^{\alpha\beta}~,~\epsilon^{12}=1.
\label{bt}
\end{equation}
where the last expression is a simple choice for $\omega ^{\alpha \beta}_
{\mu\nu}$.
The SCCs $\Theta ^\mu_\alpha$ are modified to $\tilde\Theta ^\mu_\alpha$ such 
that they
become FCC,
\begin{equation}
\{\tilde\Theta ^\mu_\alpha (q,\phi ) ,\tilde\Theta ^\nu_\beta (q,\phi )\}=0
~~;~~\tilde\Theta ^\mu_\alpha (q,\phi )=\Theta ^\mu_\alpha (q)+
\Sigma _{n=1}^\infty \tilde\Theta ^{\mu(n)} _\alpha (q,\phi )~~;~~
\tilde\Theta ^{\mu(n)}\approx O(\phi ^n),
\label{b1}
\end{equation}
with $q$ denoting the original degrees of freedom. Let us introduce the gauge 
invariant
variables $\tilde f(q)$ \cite{bt}
 corresponding to each $f(q)$, so that $\{\tilde f(q),\tilde \Theta _\alpha 
^\mu \}=0$
\begin{equation}
\tilde f(q,\phi )\equiv f(\tilde q)
=f(q)+\Sigma _{n=1}^\infty \tilde f(q,\phi )^{(n)},
\label{b7}
\end{equation}
which further satisfy \cite{bt},
\begin{equation}
\{q_1 ,q_2\}_{DB}=q_3~\rightarrow
\{\tilde q_1 ,\tilde q_2\}=\tilde q_3~~,~~\tilde 0=0.
\label{til}
\end{equation}
It is now clear that our target is to obtain $\tilde x_\mu$ for $x_\mu$. 
Explicit
expressions for $\tilde\Theta ^{\mu(n)}$ and $\tilde f^{(n)}$ are derived in 
\cite{bt}.

Before we plunge into the BT analysis, the reducibility of the SCCs
$\Theta ^\mu_1 ~(i.e. ~P_\mu \Theta ^\mu_1$=0)  \cite{sg1,sg3} is to be 
removed \cite{bn}
by introducing a canonical pair of auxiliary
variables $\phi$ and $\pi$ that satisfy $\{\phi ,\pi \}=1$ and PB commute
with the rest of the physical variables. The modified SCCs that appear in the 
subsequent
BT analysis are as shown below:
\begin{equation}
\Theta _1^\mu\equiv S^{\mu\nu}P_\nu +k_1P^\mu \pi ~~~;
~~~\Theta _2^\mu
\equiv (\Lambda ^{0\mu}-\frac {P^\mu}{M})+k_2 (\Lambda
^{0\mu}+\frac {P^\mu}{M})\phi ~, \label{150}
\end{equation}
where $k_1$ and $k_2$ denote two arbitrary parameters. Since the computations 
are
exhaustively done in \cite{sg3} they are not repeated here. The results are 
the following:
\begin{equation}
\tilde x_\mu =x_\mu +
[S_{\nu\mu}+2k_1\pi g_{\nu\mu}](\phi ^1)^\nu
+{{\cal R}_1}_{\mu\nu}(\phi ^2)^\nu +higher-\phi -terms~,
\label{x}
\end{equation}
\begin{equation}
\{\tilde x_\mu ,\tilde x_\nu \}=-\frac{\tilde S_{\mu\nu}}{M^2}~~,~~
\tilde S_{\mu\nu} =S_{\mu\nu}+{{\cal R}_2}_{(\alpha )\mu\nu\lambda}\phi ^
{(\alpha)\lambda }
+higher-\phi -terms~,
\label{s}
\end{equation}
where the expressions for ${\cal R}$ are straightforward to obtain
\cite{sg3} but are not needed in the present order of analysis. Only it 
should be remembered
that the ${\cal R}_1$-term in (\ref{x}) is responsible for the $(\phi 
^\alpha )^\mu$-free
term $-S_{\mu\nu}/(M^2)$ in the $\{\tilde x_\mu ,\tilde x_\nu \}$ bracket in 
(\ref{s}).
Thus the
problem that we had set out to solve has been addressed
successfully in (\ref{x}), which expresses the NC $\tilde x_\mu $
in terms of ordinary $x_\mu $ and other variables \cite{sg3}.
\vskip 1cm
\begin{center}
{\bf {V. BATALIN-TYUTIN EXTENSION IN A GAUGE $\rightarrow$ \\
THE SEIBERG-WITTEN MAP}}
\end{center}
\vskip .5cm
Now comes the crucial part of
identification of the present map with the SWM \cite{sw}. This means in 
particular that we
have to connect (\ref{x}) to (\ref{5}), since as we have shown before, (\ref
{5}) is capable of
generating the SWM \cite{sw}. We
exploit the freedom of choosing  gauges according to our
convenience, since in the BT extended space $\tilde \Theta _\alpha
^\mu $ are FCCs.  For instance, the so called unitary gauge, $\phi
_1^\mu=0,\phi _2^\mu=0$, trivially converts the system back to its
original form  before the BT extension. Let us choose the
following non-trivial gauge,
\begin{equation}
\phi _1^\mu=\frac{M^2}{2}A^\mu (x)~~,~~\phi _2^\mu=0,
\label{gauge}
\end{equation}
where $A^\mu (x)$ is some function of $x_\mu$, to be identified with the 
gauge field.
Let us also work with terms linear in $A^\mu (x)$. Identifying $\tilde S_
{\mu\nu}/(M)^2=
\theta _{\mu\nu}$ we end up with the cherished mapping,
\begin{equation}
\tilde x_\mu =x_\mu -\frac{1}{2}\theta _{\mu\nu}A^\nu (x) +higher-A(x)-terms 
~,
\label{b115}
\end{equation}
\begin{equation}
\{\tilde x_\mu ,\tilde x_\nu \}_{DB}
=\theta _{\mu\nu} +higher-A(x)-terms~ .
\label{db3}
\end{equation}
Note that in the above relations (\ref{b115},\ref{db3}), we have dropped the 
terms
containing $k_1$, an arbitrary parameter \cite{sg3}, considering it to be 
very small.
Also in (\ref{db3}) Dirac Bracket reappears since the system is gauge fixed 
and hence has SCCs.
This constitutes the second part of our result.

Finally, two points are to be noted. Firstly, the
non-commutativity present here does {\it not} break Lorentz
invariance since there appear no constant parameter
with non-trivial Lorentz index to start with. The violation will
appear only in the identification of $\tilde S_{\mu\nu}$ with
(constant) $\theta _{\mu\nu}$. Secondly, (\ref{x}) truly expresses
the NC space-time $\tilde x_\mu$ in terms of ordinary space-time
$x_\mu$. But $x_\mu$ becomes NC owing to the Dirac brackets
induced by the particular gauge that we fixed in order to reduce
our results to the SWM. Obviously, in general, there is no need to
fix this particular gauge. This refers to the comment below (\ref{7}).

\vskip 1cm
\begin{center}
{\bf {VI. CONCLUSIONS AND FUTURE PROSPECTS}}
\end{center}
\vskip .5cm

In conclusion, let us summarize our work.  We have shown that the
(abelian $O(\theta )$) Seiberg-Witten map can be viewed as a
co-ordinate transformation, albeit with field dependent
parameters. The duality concept between gauge orbits in
noncommutative and ordinary space-times, which was crucial in the
original derivation \cite{sw}, is not applied here. It has been
explicitly demonstrated that a noncommutative space-time sector
can be constructed in the Batalin-Tyutin extension of the
relativistic spinning particle model \cite{sg3}. Finally, the
above mentioned transformation and subsequently a direct
connection with the Seiberg-Witten map is also generated in this
model. The present work reveals that noncommutative space-time is
endowed with spin degrees of freedom, as compared to the ordinary
space-time \cite{gh2}.

As a future work, we plan to develop the quantum theory of the
noncommutative spacetime model proposed in \cite{gh2} whose
particle content is analogous to the spinning particles considered
here. This theory has a classical conformal invariance, the fate
of which will be studied upon quantization. Also, it would be
interesting to investigate how the constraints induce the
noncommutativity in spacetime coordinates in an operator product
expansion approach.

\vskip 1cm Acknowledgement: It is a pleasure to thank Professor
R.Jackiw for helpful correspondence.


\newpage
\begin{thebibliography}{99}
\bibitem {sn} H.S.Snyder, Phys.Rev. 71 38(1947).
\bibitem{lor} S.Doplicher, K.Fredengagen and J.E.Roberts, Phys.Lett. B331 39
(1994);
C.E.Carlson,
C.D.Carrone and N.Zobin, {\it Non-commutative gauge theory without Lorentz 
violation}, Phys.Rev. D66 (2002) 075001 (HEP-TH/0206035).
\bibitem {sg3} S.Ghosh, Phys.Rev. D66 045031(2002).
\bibitem {mor} H.Kase, K.Morita, Y.Okumura and E.Umezawa, {\it Lorentz-
Invariant Non-Commutative Space-Time Based On DFR Algebra}, HEP-TH/0212127.
\bibitem {gh2} S.Ghosh, {\it {Modelling a noncommutative  two-brane}}, hep-
th/0212123.
\bibitem {rev}For a review see M.R.Douglas and N.A.Nekrasov, Rev.Mod.Phys. 73 
977(2002) ( arXiv: HEP-TH/
0106048); R.J.Szabo, HEP-TH/0109162; A.Connes, {\it Noncommutative Geometry}, 
Academic Press, 1994.
\bibitem{sw}N.Seiberg and E.Witten, JHEP 09(1999)032.
\bibitem {bbg}R.Banerjee, B.Chakraborty and S.Ghosh, Phys.Lett. B537 340
(2002).
\bibitem {of} Y.Okawa and H.Ooguri, Phys.Rev. D64 046009(2001); S.Fidanza, 
JHEP 0206 016(2002).
\bibitem {phen} X.Calmet et. al., HEP-PH/0111115; H. Bozkaya, P. Fischer, H. 
Grosse, M. Pitschmann, V. Putz, M. Schweda, R. Wulkenhaar,
     {\it Space/time noncommutative field theories and causality}, HEP-
TH/0209253;
         M.Chaichian, K.Nishijima and A.Tureanu, {\it Spin-Statistics and CPT 
Theorems in Noncommutative Field Theory}, HEP-TH/0209008; T.Tamaki, T.Harada, 
U.Miyamoto and T.Torii, {\it Particle velocity in noncommutative space-time},
      Phys.Rev. D66 (2002) 105003 (gr-qc/0208002); J.L.Cortes, J.Gamboa, 
F.Mendez,  {\it Noncommutativity in Field Space and Lorentz Invariance 
Violation}, J.M.Carmona,  HEP-TH/0207158.
\bibitem {sb} R.Banerjee and S.Ghosh, Phys.Lett. B533 162(2002).
\bibitem {ard} J.M.Gracia-Bondia and C.P.Martin, Phys.Lett. B479 321(2000); 
L.Bonora, M.Schnabl and A.Tomasiello, Phys.Lett. B485 311(2000); F.Ardalan 
and N.Sadooghi, Int.J.Mod.Phys. A16 3157(2001).
\bibitem {ss} S.Ghosh, Phys.Lett. B 558 245 (2003) (hep-th/0210107); 
M.B.Cantcheff and P.Minces, Phys.Lett. B557 283(2003) (hep-th/0212031); 
O.F.Dayi, Phys.Lett. B560 239(2003).
\bibitem{ds}S.Deser, R.Jackiw and S.Templeton, Phys.Rev.Lett. 48(1982)975; 
Ann.Phys. 140 (1982)372; S.Deser and R.Jackiw, Phys.Lett. 139B(1984)371.
\bibitem{f}E.Fradkin and F.A.Schaposnik, Phys.Lett. 338B(1994)253; 
R.Banerjee, Phys.Lett.   358B (1995)297.
\bibitem {gh} S.Ghosh, Phys.Lett. B563 112(2003).
\bibitem{d} P.A.M.Dirac, {\it Lectures on Quantum Mechanics} (Yeshiva
University Press, New York, 1964).
\bibitem{bt}I.A.Batalin and I.V.Tyutin, Int.J.Mod.Phys. A6 3255(1991).
\bibitem {sg1} S.Ghosh, Phys.Lett. B338 235(1994); J.Math.Phys. 42 5202
(2001). For the original works see, A.J.Hanson and T.Regge,
Ann.Phys. (N.Y.) 87 498(1974);
see also A.J.Hanson, T.Regge and C.Teitelboim, {\it Constrained
Hamiltonian System}, Roma, Accademia Nazionale Dei  Lincei, (1976).
\bibitem {corn}L.Cornalba, {\it D-brane physics and noncommutative Yang-Mills 
theory},
 HEP-TH/9909081; A.A.Bichl et. al., {\it Noncommutative Lorentz symmetry and 
the origin
 of the Seiberg-Witten map}, HEP-TH/0108045;
 B.Jurco and P.Schupp, Euro.Phys.J. C14 367(2000); B.Jurco and P.Schupp and 
J.Wess, Nucl.Phys. B584 784(2001); H.Liu, Nucl.Phys. B614 305(2001);
  R.Jackiw, S.-Y.Pi and A.P.Polychronakos,
 {\it Noncommutating gauge fields as a Lagrange fluid}, Phys.Rev.Lett. 88 
111603(2002)
 (HEP-TH/0206014).
\bibitem{j} Field dependent transformations have appeared in {\it eg}
R.Jackiw, Phys.Rev.Lett. 41, 1635 (1978); R. Jackiw and S.-Y. Pi,
HEP-TH/0111122.
\bibitem {bn}R.Banerjee and J.Barcelos-Neto, Ann.Phys. 265 134(1998).

\end{thebibliography}
\end{document}